\documentclass[sigplan]{acmart}
\renewcommand\footnotetextcopyrightpermission[1]{}
\usepackage{url}
\usepackage{hyperref}
\usepackage{cleveref}
\usepackage{amsmath}
\usepackage{pifont}
\usepackage{makecell}
\usepackage{multicol}
\usepackage{tikz}
\usepackage{multirow}
\usepackage{stfloats}

\usepackage[skins]{tcolorbox}

\usepackage{algorithm}
\usepackage[noend]{algpseudocode}
\usepackage[cppcommentstyle,continuouslinenumbers]{smartalgorithmic}

\usepackage{caption}
\usepackage{subcaption}
\usepackage{xspace}

\usepackage[inline]{enumitem}

\makeatletter
\let\@authorsaddresses\@empty
\makeatother


\definecolor{mediumred}{RGB}{200, 0, 0}
\definecolor{mediumgreen}{RGB}{0,150, 0}
\definecolor{mediumyellow}{RGB}{220,180,0}
\mathchardef\mhyphen="2D

\newcommand{\weakerNarwhal}{{Rorqual}\xspace}

\newcommand{\parties}[0]{\mathcal{P}}
\newcommand{\certificate}[0]{\mathcal{C}}

\algnewcommand{\WWhile}[1]{\algorithmicwhile\ #1 \textbf{do}}

\newcommand{\point}[1]{\subparagraph*{\textbf{#1.}}}
\newcommand{\Message}[2]{\ensuremath{\langle}#1:~#2\ensuremath{\rangle}}

\NewDeclarationType{Protocol}{\textsc}
\NewDeclarationType{Function}{\textrm}
\NewDeclarationType{Event}{\mathit}
\NewDeclarationType{Constant}{\mathbf}
\NewDeclarationType{MessageType}{\mathbf}

\DeclareGlobalConstant{true}
\DeclareGlobalConstant{false}

\DeclareGlobalFunction{send}
\DeclareGlobalFunction{sign}
\DeclareGlobalFunction{judge}

\DeclareGlobalEvent{deliver}
\DeclareGlobalEvent{convict}

\DeclareGlobalMessageType[mShare]{Share}
\DeclareGlobalMessageType[mVertex]{Vertex}
\DeclareGlobalMessageType[mVote]{Vote}
\DeclareGlobalMessageType[mKey]{Key}
\DeclareGlobalMessageType[mEcho]{Echo}
\DeclareGlobalMessageType[mRelay]{Relay}
\DeclareGlobalMessageType[mAck]{Ack}
\DeclareGlobalMessageType[mTimeout]{Delay}
\DeclareGlobalMessageType[mCertificate]{Certificate}
\DeclareGlobalMessageType[mRequest]{Request}

\algblockdefx[Function]{Function}{EndFunction}[2]{\textbf{function} #1(#2)}{}
\algtext*{EndFunction}  

\algblockdefx[Operation]{Operation}{EndOperation}[2]{\textbf{operation} #1(#2)}{}
\algtext*{EndOperation}  

\algblockdefx[Struct]{Struct}{EndStruct}[1]{\textbf{struct} #1:}{}
\algtext*{EndStruct}  

\algblockdefx[Upon]{Upon}{EndUpon}[1]{\textbf{upon} #1}{}
\algtext*{EndUpon}  

\algblockdefx[SW]{SW}{EndSW}[0]{\textbf{in SW:}}{}
\algtext*{EndSW}  

\algblockdefx[LocalVariables]{LocalVariables}{EndLocalVariables}[0]{\textbf{local variables:}}{}
\algtext*{EndLocalVariables}

\newtcolorbox{myframe}[2][]{%
  enhanced,colback=white,colframe=black,coltitle=black,
  sharp corners, left=2pt, right=2pt, boxrule=0.4pt,
  fonttitle=\scshape,
  attach boxed title to top left={yshift=-0.3\baselineskip-0.4pt,xshift=2mm},
  boxed title style={tile,size=minimal,left=0.5mm,right=0.5mm,
    colback=white,before upper=\strut},
  title=#2,#1
}

\title{Rorqual: Speeding up Narwhal with TEEs}

\author{Luciano Freitas}
\affiliation{%
  \institution{Télécom Paris, Institut Polytechnique de Paris\\Matter Labs}
  \country{}
}

\author{Shashank Motepalli}
\affiliation{%
  \institution{University of Toronto\\Matter Labs}
  \country{}
}

\author{Matej Pavlovic}
\affiliation{%
  \institution{Matter Labs}
  \country{}
}

\author{Benjamin Livshits}
\affiliation{%
  \institution{Matter Labs}
  \country{}
}

\date{June 2024}

\begin{abstract}
In this paper, we introduce Rorqual, a protocol designed to enhance the performance of the Narwhal Mempool by integrating Trusted Execution Environments (TEEs). Both Narwhal and Roqual are protocols based on a Directed Acyclic Graph (DAG). Compared to Narwhal, Rorqual achieves significant reductions in latency and increases throughput by streamlining the steps required to include a vertex in the DAG. The use of TEEs also reduces the communication complexity of the protocol while maintaining low computational costs. Additionally, Rorqual enhances privacy features, reducing the potential for Miner Extractable Value (MEV) attacks without the need for additional communication steps. Our protocol is resilient to known TEE vulnerabilities and can function in heterogeneous networks with both TEE and non-TEE nodes. Through rigorous analysis, we demonstrate the protocol's robustness under both normal and adversarial conditions, highlighting its improvements in throughput, latency, and security.
\end{abstract}

\begin{document}
\sloppy
\maketitle

\section{Introduction}

\begin{table}[tb]
\label{tab:complexity}
\centering
\renewcommand{\arraystretch}{1.2}
\caption{Complexity comparison of different Narwhal implementations}
\resizebox{\columnwidth}{!}{%
\begin{tabular}{|l|c|c|c|c|c|}
    \hline
    \multicolumn{1}{|c|}{\multirow{2}{*}{\bf \thead{Narwhal \\ variant}}}
    & \multirow{2}{*}{\bf \thead{Computational \\ cost}}
    & \multicolumn{2}{c|}{\bf \thead{Cerficate \\ latency}}
    & \multicolumn{2}{c|}{\bf \thead{Payload \\ latency}} \\
    \cline{3-6}
    & & \thead{Good} & \thead{Bad} & \thead{Good} & \thead{Bad} \\
    \hline \hline
    Bracha
    & \textcolor{mediumgreen}{Negligible}
    & \textcolor{black}{--} 
    & \textcolor{black}{--} 
    & \textcolor{mediumred}{$2\Delta$} 
    & \textcolor{mediumyellow}{$4\Delta$} \\
    \hline
    Pull
    & \textcolor{mediumgreen}{Negligible}
    & \textcolor{mediumred}{$2\Delta$} 
    & \textcolor{mediumred}{$3\Delta$} 
    & \textcolor{mediumred}{$2\Delta$} 
    & \textcolor{mediumred}{$5\Delta$} \\
    \hline
    Semi-Avid
    & \textcolor{mediumred}{\makecell{High}}
    & \textcolor{mediumred}{$2\Delta$} 
    & \textcolor{mediumred}{$3\Delta$} 
    & \textcolor{mediumred}{$2\Delta$} 
    & \textcolor{mediumred}{$5\Delta$} \\
    \hline\hline
    \textbf{\weakerNarwhal}
    & \textcolor{mediumyellow}{\bf \makecell{Low}}
    & \textcolor{mediumgreen}{\bf $\mathbf{1\Delta}$} 
    & \textcolor{mediumgreen}{\bf $\mathbf{1\Delta}$} 
    & \textcolor{mediumgreen}{\bf $\mathbf{1\Delta}$} 
    & \textcolor{mediumgreen}{\bf $\mathbf{1\Delta}$} \\
    \hline
\end{tabular}
}
\end{table}

In this paper, we propose \weakerNarwhal, a solution combining the best properties of Trusted Execution Environments (TEEs) and the Narwhal Mempool, which forms the basis of several DAG consensus protocols that have soared in development in recent years~\cite{allYouNeedIsDAG,bullshark,danezis2022narwhal,shoal}.


Among the results we obtain, we highlight a significant reduction of latency and a throughput increase by simplifying the steps necessary to include a vertex in the DAG shown in \cref{tab:complexity}. The use of TEE also allows us to decrease the communication complexity of the protocol without incurring hefty computational costs, which occur in alternative solutions\cite{nazirkhanova2022information}.
Our gains can be combined with any improvement made on the DAG processing side, such as the Shoal mechanism~\cite{shoal}. This modification is implemented on top of DAG-based consensus protocols and has already reduced the latency of the Bullshark consensus by more than 20\%.

The use of TEEs not only reduces the latency of the protocol but also brings important privacy features that can play a crucial role in reducing MEV attacks in blockchains. When compared to other MEV-preventing techniques that have been proposed for DAG protocols, which normally require a commitment to blind data and then the opening of secrets associated with published blocks, it eliminates the necessity of this extra communication and opens up space in blocks for the prime payload: client transactions.

We analyze the design of the protocol in light of known TEE attacks and highlight the special care we took to minimize the attack surface of our protocol given the current technology limitations. Due to the small amount of the code that we need to run inside a TEE Safe World (SW), we encourage the use of a heterogeneous network based on technologies provided by different manufacturers.

\subsection{Contributions}
\label{sec:contributions}

\weakerNarwhal, our proposed protocol, offers the following contributions consisting mainly of modifying the broadcast primitive used combined with an accountability mechanism.
\begin{itemize}
    \item \textbf{Latency.} The modification in the proposed communication mechanism allows vertices to be added in the DAG in a single message delay in the best case instead of the currently required two message delays, halving the latency. (\Cref{th:good-case-latency})
    
    \item \textbf{Graceful degradation under faults.} We demonstrate that under attack, \weakerNarwhal degrades in terms of performance more gracefully than vanilla Narwhal. That is to say, the latency of the protocol stays low under an adversarial scheduler, contrary to the original Narwhal. (\Cref{th:bad-case-latency}) 

    \item \textbf{Accountability.} The typical misbehavior of peers in distributed systems involve sending conflicting messages to other peers, which is prevented in our system by the use of TEEs. Other misbehaviors, such as dropping messages sent by the protocol, are eventually detected and attributed to the offending peer.
    (\Cref{th:accountability})
    
    
\end{itemize}

It has already been known that TEE allows for \emph{consistent} broadcast in a single message delay. Our main insight in this paper is how to achieve \emph{reliable} broadcast in the context of the Narwhal protocol.

\subsection{Paper Organization}
The rest of this paper is organized as follows.
In Section~\ref{sec:problem} we formally specify the problem of sequencing in the context of DAG-based mempools;  we also define all the properties that are guaranteed by TEEs and how they are currently given by hardware providers. We also present the attack model we use in \weakerNarwhal. 
In Section~\ref{sec:protocol} we give a full description of our protocol, highlighting the parts that are run inside the TEE and their place in the larger Narwhal algorithm. 
In Section~\ref{sec:proofs} we provide proofs of correctness and complexity of our protocol, including the communication complexity bounds for different peers in \weakerNarwhal, as well as the number of message delays necessary for the formation of the DAG in both normal operation under synchrony and asynchronous operation under attack. 
In Section~\ref{sec:related} we position our paper and contributions with respect to the state of the art of consensus and L2 sequencers. 
Finally, Section~\ref{sec:conclusions} concludes.

\section{Definitions}
\label{sec:problem}

\subsection{System Model}
\label{sec:model}

We examine a system consisting of $n \ge 3f+1$ peers, each having access to Trusted Execution Environment (TEE) hardware. Among these peers, up to $f$ may behave maliciously, running compromised operating systems. Nevertheless, these peers execute the designated protocol within the Secure World (SW) enclave. It is worth noting that these malicious entities may also run arbitrary software on the same machine in an attempt to disrupt the protocol's normal execution flow and compromise its integrity.

Participants communicate via \emph{reliable channels}, ensuring the eventual delivery of messages between non-faulty peers. We adopt a \emph{partially synchronous} communication model. In this model, a Global Stabilization Time (GST) exists such that, beyond this unspecified duration, all messages are delivered within a known time bound of $\Delta$. We also define the actual upper bound for message delivery $\delta$, which remains unknown to all entities.

\point{TEE Properties}
In this paper, we make use of four main properties guaranteed by TEEs:

\begin{enumerate}
    \item \textbf{Enclaved execution:} Enclave memory cannot be read or written from outside the enclave, at the same time preventing tampering with execution flow and information leakage;
    \item \textbf{Unbiased randomness:} Enclaves have access to hardware-assisted unbiased randomness generation;
    \item \textbf{Remote attestation:} Any peer can verify that a protocol is running inside an enclave and also identify the code being executed;
    \item \textbf{Trusted elapsed time:} It is possible to obtain the elapsed time since a reference point.
\end{enumerate}

\point{TEE vulnerabilities}
To uphold the promises made by TEE hardware, it is imperative to maintain their integrity. Numerous attacks targeting various TEE technologies have been documented, along with potential countermeasures. We refer readers to the survey \cite{sgxAttacksSurvey} focusing on Intel SGX as a foundational resource. This paper adopts this analysis to classify TEE vulnerabilities into six categories:

\begin{multicols}{2}
\begin{enumerate}
    \item Address Translation; 
    \item CPU cache;
    \item DRAM;
    \item Branch prediction;
    \item Enclave software;
    \item Hardware.
\end{enumerate}
\end{multicols}
Attacks on the first three components manipulate the states of the machine components to extract information from the victim program, often using timing analysis based on access patterns. Branch prediction attacks aim to discern the algorithm's execution flow, specifically regarding conditional jumps. The last two types exploit either software bugs in SGX or vulnerabilities at the hardware level.

While attacks of the latter two kinds fall outside our purview, they are less common and can generally be rectified through updates. Consequently, our protocol design presumes a bug-free TEE and tamper-resistant hardware.
We highlight the fact that persisting data produced in the TEE is susceptible to its own vector of attacks, but these are irrelevant to our study since we only rely on volatile information.

\point{Our attack model}
We operate under the assumption that an adversary has exhaustive knowledge of the memory addresses, their usages, and the outcomes of all control flow evaluations in our code. This information can be exploited as side channels. Additionally, the adversary can intercept, delay, or even replay messages to and from the compromised enclave process.

To safeguard the data integrity of our protocol in line with what is suggested by Randmets~\cite{randmets2021overview}, we mandate that the executed codes be formulated in a ``circuit-like'' manner for confidential information, ensuring consistent memory access patterns and execution flows regardless of the data involved. For other attack vectors, our protocol provides resilience by design.

\subsection{Properties of the Narwhal Mempool}
\label{sec:properties}

This section delineates the properties of the Narwhal mempool, which serves as a communication abstraction for both the dispersal and retrieval of transaction blocks. Participants generate these blocks by accumulating transactions from clients and organizing them based on references to previously received blocks. The main goal of this section is to formalise all properties that must be kept by Rorqual.

In a specific round $r$, a peer $p_i$ disseminates a vertex $v$ which encapsulates a block payload $b$ and links to a set of vertices called $parents$ and denoted by $S$. The vertex $v$ is referred to by its digest $d$, obtained via a $digest()$ function. The mempool's interface allows $p_i$ to invoke the procedure Disperse$(v,r)$, subsequently generating a certificate of retrievability $\certificate$. We say that the dispersal is complete once the first correct peer generates a certificate of retrievability. Other peers can then invoke Verify$(i, r, d, \certificate)$, affirming the successful dispersal of $v$. Eventually, peers may generate an event Receive$(v, i, r)$, thereby learning the value of $v$.

The structure of the mempool defines a \emph{happens-before} $(\rightarrow)$ relation as follows:

\begin{itemize}
    \item If $v_1.digest() \in v_2.S$, then $v_1 \rightarrow v_2$;
    \item If $v_1 \rightarrow v_2$ and $v_2 \rightarrow v_3$, then $v_1 \rightarrow v_3$.
\end{itemize}
Parties can make use of $\text{read\_causal}(v)$ to receive a set of vertices $V$, each of which happens-before $v$, completing the list of all behaviors associated with the Narwhal mempool, which adheres to the following properties:

\begin{itemize}
    \item \textbf{Block-Availability:} If a correct peer $p_i$ receives a certificate $\certificate$ accompanied by a peer id $j$, round $r$, and digest $d$ s.t. Verify$(j, r, d, \certificate) = \text{true}$, then $p_i$ will eventually generate the event Receive$(v, i, r)$ s.t. $v.digest() = d$.
    
    \item \textbf{Consistency:} If two correct peers $p_i$ and $p_j$ generate events Receive$(v', i, r)$ and Receive$(v'', i, r)$, respectively, then $v' = v''$.
    
    \item \textbf{Containment:} For $V = \text{read\_causal}(v)$, any $v' \in V$ satisfies \( \text{read\_causal}(v') \subseteq V \).
    
    \item \textbf{2/3 Causality:} For $V = \text{read\_causal}(v)$, $V$ contains at least \( \frac{n-f}{n} \) of the blocks successfully disseminated prior to $v$, which for the maximal resilience of the protocol corresponds to $2/3$ of the blocks.
    
    \item \textbf{1/2 Chain Quality:} Within $V = \text{read\_causal}(v)$, at least \( \frac{n-2f}{n-f} \) of the blocks are generated by correct peers, which, for the maximal resilience of the protocol, corresponds to $1/2$ of the blocks.
\end{itemize}

We shall call \emph{\weakerNarwhal} a protocol that provides, with high probability%
\footnote{The negligible probability of violating \weakerNarwhal's properties corresponds to the probability of two or more peers locally generating the same random cryptographic keys. Otherwise, \weakerNarwhal guarantees its properties deterministically.},
\emph{consistency, containment, 2/3 causality and 1/2 chain quality}, but replaces \emph{block-availability} with the following property:
\begin{itemize}
    \item \textbf{Integral ancestry:} If $v'$ is dispersed by a correct peer, then $\forall v'': v'' \rightarrow v'$ all correct peers eventually generate the event Receive$(v'', v''.source, v''.round)$.
\end{itemize}

\subsection{Original Narwhal Broadcast: Pull Broadcast}
\label{sec:narwhal}

The original Narwhal paper~\cite{danezis2022narwhal} specifies a reliable broadcast strategy that the authors called quorum-based broadcast with a pull strategy. In order to make this protocol easy to refer to, we shall call it simply ``Pull Broadcast'', referring to its most notable feature.
This approach offers two notable advantages: rapid certificate deliveries under synchronous network conditions and the cessation of re-transmissions for all undelivered messages that remain pending once the sender advances to the next round.
While the authors articulate their protocol descriptively, they fall short of presenting a comprehensive pseudo-code.
To address this gap, we provide it in Algorithm~\ref{alg:qbcast}, derived from the official implementation available on the project's repository~\cite{narwhalGit}.

\begin{algorithm}[!htb]
\begin{smartalgorithmic}[1]
    \footnotesize
        \Procedure{disperse}{$v, r$}
            \State $\sigma \gets sign(v, i, r)$ to all
            \State send $\Message{\mVertex}{v, i, r, \sigma}$ to all
        \EndProcedure

        \Upon{receiving $vertex = \Message{\mVertex}{v, j, r, \sigma_h}$}
            \State check $\sigma_h$
            \State save $\{v, j, r, \sigma_h\}$ to memory
            \If{$\mVertex$ comes from $p_j$}
                \State $\sigma_v \gets sign(vertex.digest(), i, j, r)$
                \State send $\Message{\mVote}{vertex.digest(), i, j, r, \sigma_v}$ to $p_j$
            \EndIf
        \EndUpon

        \Upon{receiving $\forall p_j \in$ a quorum $Q$ $\Message{\mVote}{d, j, i, r, \sigma_j}$}
            \State $\forall p_j \in Q:$ check $\sigma_j$
            \State $\certificate_k^r \gets \bigcup_{p_j \in Q} \{\sigma_j\}$
        \EndUpon

        \Upon{generating $\certificate_j^r$ for a digest $d$ or receiving it in some message}
            \While{$\{v, j, r, \sigma\}$ not in memory}
                \State Send $\Message{\mRequest}{(\certificate_j^r, d)}$ to random peers
            \EndWhile
            \vspace{-0.5em}
            \State Generate event Receive$(v, j, r)$
        \EndUpon

        \Upon{receiving $\Message{\mRequest}{(\certificate_k^r, d)}$} from $p_j$
            \If{$\{v, k, r\}$ in memory matching the request}
                \State send $\Message{\mVertex}{v, k, r, \sigma}$
            \EndIf
        \EndUpon
\caption{Pull broadcast code for $p_i$}
\label{alg:qbcast}
\end{smartalgorithmic}
\end{algorithm}

As we can see, to disperse a vertex for a given round, the original Narwhal implementation has the source peer pack everything into a vertex, sign it, and send it to all peers. When this message arrives at other peers, they simply echo the digest of what they receive from the source with their own signature. Once a quorum of signatures is gathered for the digest of a vertex, they are combined into a certificate of availability. It is important to note that the original implementation does not rely on threshold signatures, but rather keeps the set of signatures in their raw form. The certificates are used by different peers to advance to the next round by collecting at least $n-f$ of them, which are then included in the new vertex. At this point, different peers can come across these certificates of availability without knowing the contents of the vertex. When this happens, peers select a small sample of nodes and send them a request for the contents of the missing vertex, which can then be sent by peers that have already stored it in their memory. Since at least $n-2f \ge f+1$ correct peers must have the data, the probability of receiving the data exponentially increases with each attempt.    
\section{Implementing {\weakerNarwhal}}
\label{sec:protocol}

\begin{algorithm}[tb]
\begin{smartalgorithmic}[1]
    \footnotesize
        \State $PUB\_SK$ -- map from ID to SW signature verification public key
        \State $PUB\_NK$ -- map from ID to NW signature verification public key
        \State $nk$ -- private key for signing in NW
        \State $\Sigma$ -- map from ID to quorum of signatures
        \State $LDR$ -- map from ID to last delayed round
        \State $V$ -- map from (ID, round) to vertex (holds mock values for round $0$)
        \State $DAG$ -- map from round to vertices (holds mock values for round $0$)
        \State $H\_SHARES$ -- map from (ID, round) to (index, share)
        
        \SW
            \State $sk$ -- private key for signing in SW
            \State $round$ -- round of last dispersal, init: $0$
            \State $delay$ -- round of last delay, init: $0$
        \EndSW    
\caption{{\weakerNarwhal} local variables}
\label{alg:localvariables}
\end{smartalgorithmic}
\end{algorithm}

\subsection{Overview}
In this section, we describe in detail the application of TEEs to build our {\weakerNarwhal} Mempool. The most significant difference with respect to Narwhal is that our protocol can include a vertex in the DAG with a single message delay, leveraging the TEE to guarantee consistency without additional communication. At a high level, we execute a setup procedure that guarantees that peers are running {\weakerNarwhal} inside a TEE, and renders replay attacks infeasible. Then, the bulk of the protocol consists of sending vertices to all peers and keeping track of the behavior of other peers to build edges in the DAG that guarantee better performance. 

Throughout the protocol, we make use of the variables described in Algorithm~\ref{alg:localvariables}. We assume that these variables are persistent and that each instruction we describe is atomically executed.

\point{Notation} We shall often use the notation $multicast(m)$. It means that during a period of time $\Delta$, the peer executing this instruction sends $m$ to all peers. After that, it stops sending this message once $n-f$ peers confirm the receipt of $m$.
Another keyword we introduce is \textbf{in SW}, which indicates that the instructions enclosed by this block must be executed inside the enclave, as opposed to the other instructions that are in the NW ("Normal World").

\subsection{Setup}
A significant security challenge faced by Trusted Execution Environments~(TEEs) is the \emph{replay} attack. In such an attack, an adversary may initiate multiple instances of the same protocol or even reboot machines to exploit the non-deterministic nature of distributed protocols, arising, for example, from varying message delivery sequences, to produce conflicting states across different instances of the enclave. To mitigate this vulnerability, we leverage an unbiased randomness source within the enclave to generate unique private keys, mimicking the strategy presented in~\cite{p2ptee} for their sequence numbers. The private key is never shared outside the SW, not even to processes running on the same machine, but the public key is shared with all peers, allowing them to verify messages that must originate from routines run in the SW. We emphasize that all peers must execute the entire code without terminating the enclave process. However, crashed peers can re-enter the system by using a reconfiguration protocol as new participants.

A combination of remote attestation and one round of trip of communication (gathering a quorum) guarantees that by the end of setup, each peer has the private key inside its enclave and a copy of the matching public key is held by all other peers if $p_i$ is correct (the public key is shared using consistent broadcast). On the other hand, if $p_i$ is Byzantine, then some peers might not have a copy of its public key after setup, but, most importantly, there cannot be two valid public keys.

\begin{algorithm}[tb]
\begin{smartalgorithmic}[1]
    \footnotesize
    \SW
        \Upon{Initialization}
            \State generate random $k$ and matching $pub\_k$ \label{line:genSEQ}
            \State send $\Message{\mKey}{pub\_k}$ to all
        \EndUpon
    \EndSW

    \Upon{receiving first $\Message{\mKey}{pub\_k_j}$ from $p_j$} \label{line:ackfirst}
        \State remotely attest that $p_j$ is running {\weakerNarwhal} in TEE
        \LineComment{The following signature uses regular PKI in NW}
        \State $\sigma \gets sign(pub\_k_j)$
        \State multicast $\Message{\mEcho}{pub\_k_j, \sigma}$
    \EndUpon

    \Upon{receiving $\forall p_j \in$ a quorum $Q: \Message{\mEcho}{pub\_k_k, \sigma_j}$}
        \State $\forall p_j \in Q:$ check $\sigma_j$
        \LineComment{Drop message if any signature does not check}
        \State $PUB\_K[k] \gets pub\_k_k$
        \State $\Sigma[k] \gets {aggregate}_{p_j \in Q} \{\sigma_j\})$
    \EndUpon
\caption{{\weakerNarwhal} setup for $p_i$}
\label{alg:setup}
\end{smartalgorithmic}
\end{algorithm}
    
\subsection{Vertex Dispersal}
\label{sec:header-dispersal}

To transmit vertices, our protocol employs Reed-Solomon erasure coding to fragment each vertex into \(n\) shares. A subset of these shares, specifically \(n-2f\), can be reassembled to recover the original vertex. The rationale behind this choice is two-fold. First, in alignment with pull-broadcast principles, the goal is to discontinue channel communication without reaching all peers once a vertex has been acknowledged by \(n-f\) peers. Second, as it will become clearer once we present the complexity analysis of our algorithm, by having peers send the shares of vertices they receive, we can guarantee a better latency of our protocol without increasing the communication complexity.

\begin{algorithm}[!htb]
\begin{smartalgorithmic}[1]
    \footnotesize
    \Procedure{disperse}{$v, r$}
        \SW
            \If{$round \ge r$} \label{line:monotonicRound}
                \State \textbf{Asynchronous Exit}
            \EndIf
            
            \State $\sigma_h \gets sign~((v, i, r, delay), sk)$
            \State $hRS \gets reed\mhyphen solomon((v, i, r, delay), n, n-2f)$
            \For{$p_j \in \parties$}
                \State $\sigma_j \gets sign~((hRS[j], j, i, r), sk)$
                \State send $\Message{\mVertex}{v, i, r, delay, hRS[j], \sigma_v, \sigma_j}$ to $p_j$
            \EndFor
            \State \textbf{wait} $2\Delta$
            \If{receives less than $n-f$ $\Message{\mAck}{i,r}$}
                \State $delay \gets round$
            \EndIf
        \EndSW
    \EndProcedure

    \Upon{rec. $\Message{\mVertex}{v, j, r, delay, hRS_i, \sigma_v, \sigma_i}$} from $p_j$
        \LineComment{Pull PUB\_K[j] before continuing}
        \State check $\sigma_v$ and $\sigma_i$ with $PUB\_K[j]$ \label{line:checkSEQ}
        \State $LDR[j] \gets \max(LDR[j], delay)$
        \State multicast $\Message{\mShare}{hRS_i, j, r, \sigma_i}$
        \State send $\Message{\mAck}{j, r, \sigma_i}$ to $p_j$
        \State $V[j][r] \gets v$
    \EndUpon

    \Upon{receiving $\Message{\mShare}{hRS_j, k, r, \sigma_j}$}
        \State Check $\sigma_j$
        \State $H\mhyphen Shares[k][r][j] \gets hRS_j$
    \EndUpon

    \Upon{$|H\mhyphen Shares| \ge n - 2f$}
        \If{$V[k][r] = \bot$}
            decode $V[k][r]$
        \EndIf
    \EndUpon

    \Upon{$v = V[j][r] \neq \bot$ and $\forall v' \in v.ancestor: v' \in DAG[v'.round]$}
        \State check $v.weak\_signatures$ for $v.weak\_edges$
        \State add $v$ to DAG[r]
    \EndUpon
\caption{{\weakerNarwhal} dispersal for peer $p_i$}
\label{alg:dispersal}
\end{smartalgorithmic}
\end{algorithm}

As is common for protocols designed for partially synchronous networks, our protocol is capable of making progress when the network is asynchronous (before GST), but we optimize performance once stabilization is achieved. To do that, we use the concept of delayed vertices, as we will soon explore in more detail. For now, we observe the first of the two manners a vertex can be delayed: after GST, if $p_i$ is correct, then the vertices it sends are received by all correct peers in at most $\Delta$ time, who in turn send an ack message back in at most $2\Delta$ time from the moment the vertex was sent. Thus, the enclave starts a timer for $2\Delta$ when it sends the vertex. If $p_i$ does not receive at least $n-f$ ack messages (and uses them to stop the timer in the enclave), then the TEE marks the vertex as late, updating the $delay$ variable for the current round.

Vertices are included in the DAG with the same requirement as in the original Narwhal: it is necessary to have already included their entire causal history in the DAG. Moreover, there are two types of edges in the DAG: strong edges that link a vertex to at least $n-f$ vertices in the previous round (its parents); and weak edges, to vertices in previous rounds that had been received but not yet in the causal history of any vertex in later rounds. We impose one extra condition in the formation of weak edges: peers must have already received at least $n-f$ shares for a vertex in order to include it in weak edges, together with the signatures that prove a quorum was received.

\begin{algorithm}[!htb]
\begin{smartalgorithmic}[1]
    \footnotesize

    \Upon{receiving $v$ s.t. some $v' \in v.ancestors$ has not been received}
        \SW
            \While{$\{V[j][r] = \bot\}$}
                \State Send $\Message{\mRequest}{j,r}$ to random peers.
            \EndWhile
        \EndSW
    \EndUpon
    
    \Upon{receiving $\Message{\mRequest}{k,r}$} from $p_j$
        \State send $\Message{\mRelay}{V[k][r], k, r, \Sigma_v[k][r]}$ to $p_j$
    \EndUpon
    
    \Upon{receiving $\Message{\mRelay}{v, j, r, \sigma}$}
        \State check $\sigma$
        \State $V[j][r] \gets v$
    \EndUpon
\caption{{\weakerNarwhal} pull recover for peer $p_i$}
\label{alg:pull}
\end{smartalgorithmic}
\end{algorithm}

Similarly to the original Narwhal implementation, we implement a pull mechanism for recovering missed vertices. Vertices can be missed if a malicious peer does not send them to all peers or due to asynchrony, when correct peers' messages take too long to arrive. In both cases, this mechanism is a necessary fallback that we intend to execute as few times as possible during synchrony by a deliberate selection of the vertices to include in the causal history of their proposals. 

\subsection{Choice of Parents}

\begin{algorithm}[!htb]
\begin{smartalgorithmic}[1]
    \footnotesize    
    \Upon{sending $\mVertex$ for round $r$}
        \State \textbf{wait} for $6\Delta$
        \For{$p_j$ s.t. $\not \exists (v', j, r')$ where $r' \ge r$ in memory}
            \State multicast $(\Message{\mTimeout}{j, r})$
        \EndFor
    \EndUpon
     
    \Upon{receiving $f+1$ $\Message{\mTimeout}{j,r}$}
        \State $LDR[j] \gets \max(r, LDR[j])$
    \EndUpon    

        \Upon{forming new vertex} \label{line:startVertex}
            \State $r \gets argmax(V[i][r] \neq \bot)$
            \If{$\exists r' > r$ s.t. $DAG[r']$ has $n-f$ vertices}
                \State $r\gets r'$
            \EndIf
            \State create vertex $v_{new}$ with $v_{new} = r + 1$
            \State classify vertices $v \in DAG[r]$ in one of the following categories:
                \State \quad I $|H\mhyphen Shares[v.source][v.round]| \ge n - f$ 
                \LineComment{$rho$ can be tuned to deter misbehavior}
                \State \quad II $LDR[v.source] \le r - rho$ and $\forall v' \in v.parents: |H\mhyphen Shares[v'.source][r-1]| \ge n - f$
                \State \quad III $\forall v' \in v.parents: |H\mhyphen Shares[v'.source][r-1]| \ge n - f$
                \State \quad IV other vertices
            \If{Before $2\Delta$ since last vertex}
                \State set $v_{new}$ parents as first $n-f$ vertices in I and II, in this order
            \Else
                \State set $v_{new}$ parents as first $n-f$ vertices in I, II, III, and IV, in this order
            \EndIf
            \State $v_{new}.weak\_edges \gets \{v'' : v''$ not in $v_{new}$ path and $|H\mhyphen Shares[v''.source][v''.round]| \ge n-f\}$
            \State $v.weak\_signatures \gets \Sigma\mhyphen Shares[v''.source][v''.round]$
        \EndUpon \label{line:endVertex}
        
\caption{{\weakerNarwhal} parent choice for peer $p_i$}
\label{alg:parents}
\end{smartalgorithmic}
\end{algorithm}

As we shall later prove, if a correct peer sends a vertex for round $r$, then if the network is synchronous, all correct peers are capable of sending vertices for rounds from $r$ onwards after at most $6\Delta$ time, relying on the pull recovery mechanism we previously presented.
Thus, the second form of delay is characterized by missing this delay. After this time is up, peers executing {\weakerNarwhal} send a delay message marking all peers they have not received a vertex from for rounds advanced enough. If $f+1$ peers mark a peer $p_j$ as delayed, then all peers increase their local $LDR$ entry for $p_j$. Contrary to Narwhal, we allow peers to skip to rounds where they have already received at least $n-f$ vertices.

It is enough to guarantee protocol correctness to simply wait for the first $n-f$ vertices to be added to the DAG and use them as parents for the next vertex. However, this simple approach has the drawback of being susceptible to malicious attacks where some peers do not send their vertices to all others. This slows down the progress of correct peers that inadvertently select some of these vertices in the causal history of their proposals. This, in turn, delays the inclusion of those peers' vertices into other correct peers' DAGs. We have already laid down the foundations on how we circumvent this problem with the use of erasure cording and the definition of vertex delay (\Cref{sec:header-dispersal}), thus we categorize vertices in the four categories presented in Algorithm~\ref{alg:parents}, selecting vertices with an order that, as we shall shortly prove, minimizes the probability of slowing down the protocol.

\section{Analysis of Rorqual}
\label{sec:proofs}

This section presents a proof of correctness in Section~\ref{sec:proof}, 
an analysis of time complexity in Section~\ref{sec:time-complexity} and 
comunication complexity in 
Section~\ref{sec:comms-complexity}.

\subsection{Proof of Correctness}
\label{sec:proof}

\begin{lemma}
    For each peer $p_i$, there is a unique valid private key used for signing values in the SW. 
    \begin{proof}
        

        At initialization, every peer independently generates a pair of public and private keys in the SW, relying on their unbiased source of random number generation. These public keys are then broadcast to all peers in the NW.

        Each peer waits for acknowledgments from \( n-f \) other peers before accepting a public key, which they check to have been produced by an eclave executing {\weakerNarwhal} via the remote attestation procedure. Assume for contradiction that two correct peers accept differing keys. In that case, at least \( f+1 \) peers must have acknowledged both conflicting keys. Consequently, at least one correct peer would have to acknowledge both, contradicting the constraint at line~\ref{line:ackfirst} that allows for acknowledging only one of them.
    \end{proof}
    \label{lem:uniqueSEQ}
\end{lemma}

\begin{theorem}
    {\weakerNarwhal} guarantees \emph{consistency} with high probability.
    If two correct peers $p_i$ and $p_j$ set $V[k][r]$ to $v'$ and $v''$, respectively (this is equivalent to receiving these values), then the probability of $v' \neq v''$ is negligible.
    \begin{proof}



    Parties may receive a vertex in three ways. First, they may receive a \(\mVertex\) message, requiring a TEE SW-generated signature. This signature, due to the \emph{enclaved execution} property, obliges \( p_k \) to execute both Disperse\((v',r)\) and Disperse\((v'',r)\) in Algorithm~\ref{alg:dispersal}.

    Under standard operation, the safeguard at line~\ref{line:monotonicRound} prevents duplication. If an enclave reset occurs, or multiple instances are launched, leading the SW to repeat round numbers, because of the enclave initialization procedure, each instance ends up with different keys with high probability because of their random generation. Then according to Lemma~\ref{lem:uniqueSEQ}, only one key can produce valid signatures, ensuring only one vertex is valid. Another possibility for receiving a vertex, via the pull mechanism of Algorithm~\ref{alg:pull}, is equivalent to receiving the vertex directly from the peer that originates the vertex, as the information is simply relayed with an unforgeable signature from a previous vertex message.

    The third manner to receive a vertex is by reconstructing it upon receiving at least \( n-2f \) shares linked to a specific source and round. These shares are verified through their signatures, created by the originating TEE's private key, which by the \emph{enclaved execution} property, must not deviate from the algorithm prescribed, guaranteeing that they are correctly formed.
    \end{proof}
\end{theorem}

\begin{theorem}
    {\weakerNarwhal} guarantees \emph{integral ancestry}. If $v'$ is dispersed by a correct peer, then $\forall v'': v'' \rightarrow v'$ all correct peers eventually include $v''$ in their DAG.
    \begin{proof}
        We argue by induction that if a correct peer refers to a vertex \( v \), then it must already include \( v \)'s entire causal history. Consider \( v_k \) as the \( k \)-th vertex that a peer \( p_i \) includes in its DAG. Initially, the DAG is empty, making the property trivial. Assuming the property holds for vertices \( \{v_1, \ldots, v_k\} \), when \( p_i \) includes \( v_{k+1} \), its edges comply with lines~\ref{line:startVertex} to \ref{line:endVertex}, ensuring that links are formed only to vertices already in the DAG.

        As messages between two correct peers are eventually delivered, and vertices are retransmitted during $\Delta$ time, extending until \( n-f \) peers receive them if necessary, all correct peers will eventually receive all vertices from \( p_i \). Any missing ancestors of $v$ trigger the pull mechanism presented in Algorithm~\ref{alg:pull}, which eventually reach peers with the required vertices, including $p_i$, relaying the causal history to all correct peers.
    \end{proof}
\end{theorem}

\begin{theorem}
    {\weakerNarwhal} provides \emph{containment, 2/3 causality, 1/2 chain quality}.
    \begin{proof}
        These attributes are assured by components of the Narwhal protocol that remain unchanged in our implementation. For detailed proofs, we direct the reader to the original work by Danezis et al.~\cite{danezis2022narwhal}.
    \end{proof}
\end{theorem}

\subsection{Time Complexity}
\label{sec:time-complexity}

\subparagraph*{\bf Latency definition.} We measure the round latency of our protocol as the maximal (over all possible executions) amount of time elapsed from the time a correct peer sends a vertex until $f+1$ correct peers include this vertex in their DAG.

\begin{lemma}
    After GST, no correct peer $p_i$ ever delays vertices.
    \label{lem:nodelay}
    \begin{proof}
        In order for $p_i$ to timeout to other peers, it must either increment its value of $delay$ or have $f+1$ peers send timeout messages. After GST, once $p_i$ multicasts any vertex, it must arrive at all correct peers within $\Delta$ delay, who then send ack messages back, guaranteeing that by $2\Delta$ time from the vertex being sent $p_i$ has gathered enough responses to not increase its $delay$.

        If another correct peer $p_j$ sends a vertex $v$ at time $t$, $p_i$ can be timed out if $p_j$ does not receive any vertex for $v.round$ from $p_i$ by time $t + 6\Delta$. By the synchrony assumption, $p_i$ must receive $v$ by time $t + \Delta$, and if it does not have the full causal history of $v$, then it starts pulling the missing information by $t + 3\Delta$. Thus, after a roundtrip, placing us at time $t + 5\Delta$, $p_i$ must have enough information to propose a vertex for a round greater or equal to $v.round$, arriving at $p_j$ by time $t + 6\Delta$, thus preventing a timeout.
    \end{proof}
\end{lemma}

\begin{theorem}
    \label{th:good-case-latency}
    The good case latency of our protocol, i.e., after GST when there is no malicious behavior, is $\delta$.
    \begin{proof}
        In the case mentioned, all vertices arrive in at most $\delta$ time. Because of the combination of our synchrony assumptions and the lack of malicious behavior, all the preceding vertices must also have been delivered by this time with all peers not being timed out.
    \end{proof}

    That is effectively the low-latency property in our main contributions.
\end{theorem}

\begin{theorem}
    \label{th:accountability}
    If $t$ is the time a peer $p_i$ sends a vertex $v$ and all correct peers do not receive $v$ by $t + 2\Delta$, then all correct peers eventually increase $LDR[i]$ to at least $v.round$.
    \begin{proof}
        Suppose, by contradiction, that some correct peer $p_j$ does not receive $v$ before $t + 2\Delta$ and $LDR[v.source]$ remains less than $v.round$.
        In order for $p_i$ to not increase its local variable $delay$, it must receive $\mAck$ from at least $n-f$ peers by time $t + 2\Delta$. After GST, in order to guarantee the reception of the acknowledgments by $t+2\Delta$, at least $n-2f$ correct peers must send shares of $v$ to all peers which would reach all correct peers at the latest by $t + \Delta$, which is enough for all peers to retrieve $v$, concluding the contradiction. Thus, either $p_j$ eventually receives a new vertex that contains the updated delay, increasing $LDR[i]$, or it does not receive vertices from $p_i$ anymore. In the latter case, there will be eventually $f+1$ correct peers that will send $\mTimeout$ messages that will also increase $LDR[i]$.

    \end{proof}
    This accountability property is highlighted as one of our main contributions on detecting when peers do not send a message to enough correct peers.
\end{theorem}

\begin{theorem}
    \label{th:bad-case-latency}
    Eventually, rounds last at most $\Delta$.
    \begin{proof}
        After GST, according to lemma~\ref{lem:nodelay}, the entries in the local variables $LDR$ that assess the delays of peers stop increasing for correct peers, which allows correct peers to select parents for their vertices from categories I and II. If malicious peers delay vertices, then they stop being selected, with only peers that do not delay vertices being eventually used as vertex sources.

        Let $t_r$ be the time a correct peer $p_i$ sends its vertex $v$ for round $r$. Then, since the sources of parent vertices do not delay, at least $n-2f \ge f+1$ correct peers must receive them by $t_r + \Delta$, which is also an upper bound for receiving $v$, including it in their DAG.

        
        \end{proof}

        Since this is asymptotically the same latency of the good-case scenario as shown in \Cref{th:good-case-latency}, we say that our protocol has graceful degradation.
\end{theorem}

\subsection{Communication Complexity}
\label{sec:comms-complexity}

\begin{theorem}
    The communication complexity of our protocol is $O(n^2\kappa)$ per vertex. Where $\kappa$ is the size of a signature.
    \begin{proof}
        When sending a vertex, each peer sends to all $n$ peers messages of size $O(n\kappa)$, with the $n$ signatures present in the signatures of weak edges dominating the complexity. Strong edges can be represented by $n$ bits, indicating which of the previous vertices are being referenced. (This could be brought to $O(\kappa)$ by using threshold signatures, amounting to $O(n^2\kappa)$ communication complexity). 
        Then, all $n$ peers send to all $n$ peers a share of the vertex of size $O(\kappa)$, bringing the total amount of data exchanged in these messages also to $O(n^2 \kappa)$.
        When pulling vertices, the bulk of the cost comes from all peers potentially sending to all peers an extra copy of each vertex, tallying $O(n^2\kappa)$ data being exchanged.
        Delay messages consist purely of signatures, but can also be sent from all peers to all peers, requiring $O(n^2 \kappa)$ data exchanged.
    \end{proof}
\end{theorem}

\section{Using Rorqual with Bullshark}


\begin{theorem}
    Bullshark using {\weakerNarwhal} as its Mempool provides \emph{Safety}.
    \begin{proof}
        In order to prove that Bullshark provides this property, the original paper~\cite{bullshark} relies on two observations all the other related proofs are built upon. Specifically, for two correct peers $p_i, p_j$ with local DAGs $DAG_i$ and $DAG_j$, the following must hold:
        \begin{itemize}
            \item $\bigcup_{r>0} DAG_i[r] = \bigcup_{r>0} DAG_j[r]$. $\weakerNarwhal$ guarantees this by the fact that correct peers never stop proposing and the \emph{integral ancestry} property fulfills this requirement once peer $i$ delivers a vertex from peer $j$ for round $r$, and vice-versa.

            \item For any round $r$ with $v' \in DAG_i[r]$ and $v'' \in DAG_j[r]$ s.t. $v'.source = v''.source$, if there is a path from a vertex $v$ to $v'$ in $DAG_i$ s.t. $v.round < r$, then there is also a path from $v$ to $v''$ in $DAG_j$. These properties are simply a restating of \emph{consistency} and \emph{integral ancestry}.
        \end{itemize}
    \end{proof}
\end{theorem}
\begin{theorem}
    Bullshark using {\weakerNarwhal} as its Mempool provides \emph{Liveness}.
    \begin{proof}
        The original Bullshark uses the following observation in order to prove the liveness of their protocol:
        \begin{itemize}
            \item For every round $r$ and correct peer $p_i$, $DAG[r]$ eventually contains a vertex for every correct peer.
        \end{itemize}

    This is used in order to prove the following claims:
    \begin{itemize}
        \item Correct peers advance an unbounded number of rounds and produce an unbounded number of vertices;
        \item Every correct peer eventually has at least $n-f$ vertices for every round in the DAG;
    \end{itemize}
    Although our protocol allows for peers to skip rounds, breaking the original observation used to prove these claims, they are still valid thanks to the \emph{integral ancestry} property. First, let $r$ be the current round of the correct peer $p_i$ that advanced the most. Then, by \emph{integral ancestry}, all correct peers will eventually get the causal history of the vertex $V[i][r]$, being able to eventually reach round $r$. Thus, eventually, all peers will obtain at least the $n-f$ vertices provided by correct peers on round $r$ and be able to advance to round $r+1$.
    
    As for the second claim, it holds for round $0$ for any correct peer $p_i$ because of the initialization. Suppose this is true for a given round $r$. Since we have already established that correct peers do not stop making progress, once a correct peer $p_j$ advances to a round $r' > r + 1$, it must include at least $n-f$ vertices of round $r' - 1$ as well as their full causal history, which are eventually included by $p_i$ because of \emph{integral ancestry}.
    \end{proof}
\end{theorem}

\section{Related Work}
\label{sec:related}
\subsection{DAG Protocols}

DAG Rider~\cite{allYouNeedIsDAG} introduced the organization of proposals into a DAG, where each proposer references prior proposals, creating a structure that captures the causal history of proposals. 
This approach paved the way for the Narwhal mempool~\cite{danezis2022narwhal}, which enhanced both throughput and DAG construction efficiency. Protocols like BullShark~\cite{bullshark}, BBCA~\cite{bbca}, and Shoal~\cite{shoal} leveraged Narwhal, endowing these protocols with high throughput and low latency. These protocols rely on the standard broadcast primitives and the pull broadcast presented in \cref{alg:pull} which is the main difference from what we presented.





\subsection{TEEs}

Jia et al. \cite{p2ptee} propose an alternative approach for peer-to-peer (P2P) protocols using Trusted Execution Environments (TEEs), including reliable broadcast and a common coin. They demonstrate that their method reduces Byzantine faults to general omission faults. However, their broadcast protocol differs significantly from ours, as it requires a linear number of communication rounds based on the number of peers. This difference underscores the importance of our approach, which assumes that all peers are active broadcasters. The increased cost in their method arises from the assumption that there is only a single source inserting data into the system, whereas our protocol is designed to accommodate multiple data sources efficiently. It also highlights the contribution of our algorithm to choose parents, with an optimistic protocol that carries a great performance on the general case because of it. 

Engraft \cite{wang2022engraft} is a consensus protocol that adapts the full Raft algorithm to be executed inside the TEE. One advantage of their approach compared to ours is that they achieve resilience with $n \ge 2f+1$. However, their protocol requires an additional property of TEE that ours does not: the use of persistent memory, which increases the attack surface. Furthermore, the single-leader design in their protocol allows an adversary to periodically affect its performance, as well as not being able to use the improvements of gained by multiple sources of data, as highlighted before. In contrast, one of our contributions is the graceful degradation of our protocol, mitigating such performance impacts.

Accountability in distributed systems has also been explored in the literature \cite{raftForensics,crimeAndPunishment,EasyAsABC,AccountableLA}. All such protocols are capable of detecting misbehavior, more specifically equivocation. This is different from \weakerNarwhal where we use TEEs to prevent equivocation. We also provide accountability also against selective omission, i.e., we can eventually detect when a malicious peer sends a message to some but not all correct peers. This is a crucial step in achieving a single message-delay latency. 
Moreover, these protocols are tailored for a different system model and do not rely on TEEs or DAGs.

\section{Conclusions}
\label{sec:conclusions}
Rorqual represents a significant advancement in the optimization of DAG-based consensus protocols by leveraging the capabilities of Trusted Execution Environments. The integration of TEEs into the Narwhal Mempool protocol has resulted in notable improvements in both latency and throughput, making the protocol more efficient and secure. By addressing the communication complexity and enhancing privacy features, Rorqual effectively mitigates the risk of MEV attacks and provides a robust solution suitable for modern blockchain systems. The protocol's compatibility with existing non-TEE systems ensures a smooth transition and broader applicability. Overall, \weakerNarwhal sets a new standard for high-performance, secure, and efficient consensus mechanisms in the blockchain space.

\bibliographystyle{ACM-Reference-Format}
\bibliography{references}

\newpage

\appendix
\section{Narwhal time complexity}
\begin{theorem}
    The good case latency of Narwhal, i.e., after GST when there is no malicious behavior, is $2\delta$ to proceed to the next round as well as to deliver vertices.
    \begin{proof}
        The $n-f$ peers in the latest round will send their vertices to all peers, taking time at most $\delta$, which in turn will send their votes to all peers taking an additional time $\delta$ to reach their destination. This is enough to form the certificates of availability and proceed to the next round, as well as deliver the vertices received from the source.
    \end{proof}
\end{theorem}

\begin{theorem}
    After GST, the original Narwhal implementation has amortized worst-case of $2\Delta$ to proceed to the next round, and $5\Delta$ to deliver the vertices. 
    \begin{proof}
        Assuming that all correct peers are in the latest round, after $2\Delta$ all their vertices will have been voted by a quorum and ready to be referenced, thus malicious peers must also guarantee certificate formation by that time if they are referenced by any correct peer. Malicious peers, however, can send their values to $n-2f$ correct peers, as well as $f$ malicious peers they are in coalition with, which in turn only sends their votes to some of the initial $n-2f$ peers. Then, when the correct peers that receive the vertex from the malicious source proceed to the next round, making a new proposal that arrives at time $3\Delta$ since the beginning of the round, other correct peers need to wait for a $2\Delta$ round trip to get the vertex by the pull strategy.  
    \end{proof}
\end{theorem}

\section{Pseudo-codes for the remaining {\weakerNarwhal} modules}
\label{app:pseudocodes}

\begin{algorithm*}[b]
    \caption{Data structures for peer $p_i$}
    \label{alg:dataStructures}
    \begin{smartalgorithmic}[1]
    \footnotesize
        \Struct{$\textit{vertex } v$} 
            \Comment{The struct of a full vertex in the DAG}
        
            \State $v.\textit{round}$ - the round of $v$ in the DAG
            \State $v.\textit{source}$ - the peer that broadcast $v$
            \State $v.\textit{block}$ - a block of transactions
            \State $v.\textit{strongEdges}$ - a set of vertices in
                $v.\textit{round}-1$ that represent \emph{strong} edges, referred to by digests, \textcolor{blue}{together with the $i$-th share of these vertices}
            \State $v.\textit{weakEdges}$ - a set of vertices in rounds $<
                v.\textit{round}-1$ that represent \emph{weak} edges referred to by digests,  \textcolor{blue}{together with the $i$-th share of these vertices}
            \State \textcolor{blue}{$v.\textit{latency\_scores}$ - array with perceived latency score for each
            of the other peers}
        \EndStruct

        \State $DAG[]$ - An array of sets of vertices - initially $\varnothing$
        \State \textcolor{blue}{$V\mhyphen{Shares}[][][]$ - a 3D array of vertice shares first indexed by source, then round, then share position - initially $\bot$}
        \State \textcolor{blue}{$V[][]$ - a 2D array of vertices indexed first by source, then by round - initially $\bot$}
        \State $\textit{blocksToPropose}$ - A queue, initially empty,
            $p_i$ enqueues valid blocks of transactions from clients
        \State $\emph{round}$ - initially $1$  
        \State $buffer$ - initially $\varnothing$
        \State $\emph{wait}$ - initially $\true$
    \end{smartalgorithmic}
\end{algorithm*}

\begin{algorithm*}[!htb]
    \caption{DAG construction, protocol for process $p_i$}
    \begin{smartalgorithmic}[1]
    \footnotesize
        
        \Upon{receiving }
            \State $v \gets V[j][r]$

                \State $DAG[r] \gets DAG[r] \cup v$
           \algspace[0.1]
           
           \If{$r = round$}
                \State $\emph{w} \gets \lceil r/4 \rceil $
                \Comment{steady state wave number}
                
                \If{$r~mod~4 = 1 \wedge (\neg  \emph{wait} \vee v' \in DAG[r]: v'.source = get\_first\_steady\_vertex\_leader(\emph{w})][r])$}
                \label{line:advance1}    
                    \State \quad \quad $try\_advance\_round()$
                \EndIf
                
                \If{$ r~mod~4 = 3 \wedge (\neg  \emph{wait} \vee v' \in DAG[r]: v'.source = get\_second\_steady\_vertex\_leader(\emph{w})][r])$}
                \label{line:advance3} 
                    \State \quad \quad $try\_advance\_round()$
                \EndIf
                
                \If{$ r~mod~4 =  0 \wedge (\neg \emph{wait} \vee \exists U \subseteq DAG[r]: |U| = n-f \text{ and } \newline
                \hspace*{15mm} \forall u \in U, u.source \in \emph{steadyVoters}[ w]) \wedge 
                strong\_path(u, get\_second\_steady\_leader(\emph{w}))$}
                \label{line:advance4} 
                    \State \quad \quad $try\_advance\_round()$
                \EndIf
                
                \If{$ r~mod~4 =  2 \wedge (\neg \emph{wait}
                \vee \exists U \subseteq DAG[r]: |U| =  n-f \text{ and }  \newline
                \hspace*{15mm}
                \forall u \in U, u.source \in \emph{steadyVoters}[w]) \wedge 
                strong\_path(u, get\_first\_steady\_leader(\emph{w}))$}
                \label{line:advance2} 
                    \State \quad \quad $try\_advance\_round()$
                \EndIf
        \EndIf
        \EndUpon

        \algspace[0.3]

        \Procedure{\textit{path}}{$v,u$} \Comment{Check if exists a path consisting of strong and weak edges in the DAG}
        \State \Return exists a sequence of $k \in \mathbb{N}$,
        vertices $v_1,\dots,v_k$  s.t. $v_1 = v$, $v_k = u$, and
        \State* \quad $\forall i \in [2..k] \colon v_i \in \bigcup_{r \geq 1}
        DAG[r] \wedge v_i \in v_{i-1}.\textit{weakEdges} \cup v_{i-1}.\textit{strongEdges})$
        \EndProcedure

        \algspace[0.3]
        
        \Procedure{\textit{strong\_path}}{$v,u$} \Comment{Check if exists a path consisting of only strong edges in the DAG}
        \State \Return exists a sequence of $k \in \mathbb{N}$, vertices $v_1,\dots,v_k$  s.t. $v_1 = v $, $v_k = u$, and 
        \State* \quad $\forall i \in [2..k] \colon v_i \in \bigcup_{r \geq 1} DAG[r] \wedge v_i \in v_{i-1}.\textit{strongEdges}$
        
        \EndProcedure
        \vspace{-1em}
        
        \begin{multicols}{2}
        \Procedure{\textit{create\_new\_vertex}}{round} \label{alg:DAG:createNewVertex}
        \State \textbf{wait until}
        $\neg$\textit{blocksToPropose}.\text{empty}()
        
        \State $v.round \gets round$
        \State $v.source \gets p_i$  
        \State $v.\textit{block} \gets
        \textit{blocksToPropose}.\text{dequeue}()$
        \State \textcolor{blue}{\textbf{let} $S \subseteq \parties: \forall p_j \in S: \Sigma[j][round - 1] \neq \bot$}
        \State \textcolor{blue}{$\Sigma_S \gets \{\sigma | \forall p_j \in S: \sigma = \Sigma[j][round - 1]$}
        \State $v.\textit{strongEdges} \gets \textcolor{blue}{S, \Sigma_S}$ \label{line:createStrongE}
        \State $\textit{set\_weak\_edges}(v,\textit{round})$ \State \textbf{return} $v$
        \EndProcedure

        \algspace[0.3]
        
        \Procedure{\textit{set\_weak\_edges}}{$v, \textit{round}$} 
        \label{alg:DAG:getWeakEdges} 
        \State $v.\textit{weakEdges} \gets \{\}$
        \For{$r=\textit{round}-2$ down to 1}
        \For{\textbf{every} $u \in DAG[r]$ s.t. $\neg \textit{path}(v,u) $}
            \State $j \gets u.source$
            \State $r \gets u.round$
            \State $v.\textit{weakEdges} \gets v.\textit{weakEdges} \cup \{j, r, \textcolor{blue}{\Sigma[j][r]}\}$ \label{line:createWeakE}
        
        \EndFor
        \EndFor
        \EndProcedure

        \algspace[0.3]
        
        \Procedure{\textit{get\_fallback\_vertex\_leader}}{$w$}
            \State $p \gets \textit{choose\_leader}_i(w)$
            \State \Return $get\_vertex(p,4w-3)$ 
        \EndProcedure

        \columnbreak
        
        \Procedure{\textit{get\_first\_steady\_vertex\_leader}}{$w$}
            \State  $p \gets \textit{get\_first\_predefined\_leader}(w)$
            \State \Return $get\_vertex(p,4w-3)$ 
        \EndProcedure

        \algspace[0.3]
        
        \Procedure{\textit{get\_second\_steady\_vertex\_leader}}{$w$}
            \State  $p \gets \textit{get\_second\_predefined\_leader}(w)$
            \State \Return $get\_vertex(p,4w-1)$ 
        \EndProcedure
        
        \algspace[0.3]
        
        \Procedure{\textit{get\_vertex}}{p,r}
            \If{$\exists v\in DAG[r]$ s.t.\ $v.source = p$}
                \State \Return $v$ 
            \EndIf 
                \State \Return $\bot$ 
        \EndProcedure

        \algspace[0.3]
        
        \Upon{timeout}
            \State $\emph{wait} \gets false$
            \State $try\_advance\_round()$ 
        \EndUpon
        
        \algspace[0.3]
            
            
        
        \Procedure{try\_advance\_round}{}
        
            \If {\textcolor{blue}{$\exists S: \forall p_j \in S: \left|S\right| \geq n-f \wedge \Sigma[j][round] \neq \bot$}}
                \State $\emph{round} \gets \emph{round}+1$; \emph{start timer};
                $\emph{wait} \gets true$
                \label{line:time1}
                \State $v \gets \textit{create\_new\_vertex}(round)$ 
                \State \textcolor{blue}{$Disperse(v)$}
            \EndIf
        \EndProcedure
        \end{multicols}
        \vspace{-2em}
\end{smartalgorithmic}
\label{alg:DAG}
\end{algorithm*}

\begin{algorithm*}[!htb]
\caption{Bullshark consensus protocol}
\begin{algorithmic}[1]
 \footnotesize

    \vspace{-2em}

        \begin{multicols}{2}

        \null\vfill
            
        \LocalVariables
        
        \State $\emph{steadyVoters}[1] \gets \parties$
        \State $\emph{fallbackVoters}[1] \gets \{\}$
        \State \textbf{For every} $j > 1$, $\emph{steadyVoters}[j],\emph{fallbackVoters}[j] \gets \{\}$
        \State $ \emph{committedRound} \gets 0$
        \State $\emph{deliveredVertices} \gets \{\}$
        \State $\emph{leaderStack} \gets $ initialize empty stack

        \EndLocalVariables

        \algspace[0.3]
       
    \Procedure{try\_ordering}{$v$}
        \State $w \gets \lceil v.round / 4 \rceil$
        \State $\emph{votes} \gets v.\emph{strongEdges}$
     \If{\emph{v.round} mod 4 = 1 }
        \Comment{first round of a wave}
        
         \State \emph{determine\_peer\_vote\_type(v.source, votes, w)}
         
    \ElsIf{\emph{v.round} mod 4 = 3}
        \State \emph{try\_steady\_commit}(\emph{votes}, \emph{get\_first\_steady\_vertex\_leader(w)}, $w$)
    
    \EndIf
        
    \EndProcedure

    \algspace[0.3]
       
    \Procedure{determine\_peer\_vote\_type}{$p,votes, w$}
        \State $v_s \gets \emph{get\_second\_steady\_vertex\_leader(w-1)}$
        \State $v_f \gets \emph{get\_fallback\_vertex\_leader(w-1)}$
        
        \If{\emph{try\_steady\_commit}(\emph{votes}, $v_s$, $w-1$) $\vee$ \emph{try\_fallback\_commit}(\emph{votes, $v_f$, $w-1$})}
        
            \State $\emph{steadyVoters}[w] \gets \emph{steadyVoters}[w] \cup \{p\}$
            
        \Else
        
            \State $\emph{fallbackVoters}[w] \gets \emph{fallbackVoters}[w] \cup \{p\}$
        
        \EndIf
        
    \EndProcedure

    \algspace[0.3]
    
    \Procedure{try\_steady\_commit}{\emph{votes}, $v, w$}
        \If{$|\{ v' \in \emph{votes}: 
        v'.source \in \emph{staedyVoters[w]} \wedge  
        \hspace*{5mm} strong\_path(v',v)\}| \geq 2f+1$}
            \State $commit\_leader(v)$
            \State \Return true
        \EndIf
        \State \Return false
    \EndProcedure

    \algspace[0.3]
    
    \Procedure{try\_fallback\_commit}{\emph{votes}, $v, w$}
        \If{$|\{ v' \in \emph{votes}: 
        v'.source \in \emph{fallbackVoters[w]} \wedge  
        \hspace*{5mm} strong\_path(v',v)\}| \geq 2f+1$}
            \State $commit\_leader(v)$
            \State \Return true
        \EndIf
        \State \Return false
    \EndProcedure

    \vfill \null
    \columnbreak
    
    \Procedure{$order\_vertices()$}{}
        \While{$\neg \textit{leadersStack}.\text{isEmpty}()$} 
        \State $v \gets \textit{leadersStack}.\text{pop}()$ \label{alg:SMR:stackPop}
          \State \textit{verticesToDeliver} $\gets \{v' \in \bigcup_{r > 0}
        DAG_i[r] \mid path(v,v') \wedge v' \not\in $
        \State \hspace{30mm} $\emph{deliveredVertices}\}$
          \For{$\textbf{every} ~v' \in \textit{verticesToDeliver}$ in some deterministic
          order}
              \State \textbf{output}
              $\textit{a\_deliver}_i(v'.\textit{block},v'.\textit{round},
              v'.\textit{source})$
              \label{alg:SMR:decide}
              \State $\textit{deliveredVertices} \gets \textit{deliveredVertices} \cup \{v'\}$
          \EndFor
        \EndWhile
        \EndProcedure  

    \algspace[0.3]
    
    \Procedure{commit\_leader}{$v$}
    
        \State $\emph{leaderStack}.push(v)$
        \State $r \gets \emph{v.round} -2 $
        \Comment{There is a potential leader to commit every two rounds}
        \While{$r > \emph{committedRound}$}
        
            \State $w \gets \lceil r/4 \rceil $
            \State $\emph{ssPotentialVotes} \gets
            \{v' \in DAG_i[r+1] ~|~ strong\_path(v,v') \}$

            \If{$ r~mod~4 == 1$}
            \Comment{two potential leaders in this round}
            
               \State $v_s \gets get\_first\_steady\_vertex\_leader(w)$ 
               
               \State $v_f \gets get\_fallback\_vertex\_leader(w)$ 
               
               \State $\emph{ssVotes} \gets \{v' \in \emph{ssPotentialVotes}: v'.source \in$ \Statex \hspace{21mm} $\emph{steadyVoters}[w] \wedge  strong\_path(v',v_s) \}$ 
               
               \If{$v.round = r+2$}
               
                    \State $\emph{fbVotes} \gets \{\}$
               \Else
                    \State  $\emph{fbPotentialVotes} \gets
                    \{v' \in DAG_i[r+3] ~|~ strong\_path(v,v') \}$
                    \State $\emph{fbVotes} \gets \{v' \in \emph{fbPotentialVotes}: v'.source \in$ \Statex \hspace{26mm} $\emph{fallbackVoters}[w] \wedge  strong\_path(v',v_f) \}$

               \EndIf

            \Else
            \Comment{$ r~mod~4 == 3$}
            
                \State $v_s \gets get\_second\_steady\_vertex\_leader(w)$
                
                \State $\emph{ssVotes} \gets \{v' \in \emph{ssPotentialVotes}: v'.source \in$ \Statex \hspace{21mm} $\emph{steadyVoters}[w] \wedge  strong\_path(v',v_s) \}$ 
               
               \State $v_f \gets \bot$; $\emph{fbVotes} \gets \{\}$

            \EndIf
            
                    \If{$|\emph{ssVotes}| \geq f+1 \wedge |\emph{fbVotes}| <f+1$}
                    \label{line:indirectfirst}
            \State $leadersStack.push(v_s)$
            \label{line:stacksteady}
            \State $v \gets v_s$ 
            
        \EndIf
        
        \If{$|\emph{ssVotes}| < f+1 \wedge |\emph{fbVotes}| \geq f+1$}
         \label{line:indirectfallback}
        
            \State $leadersStack.push(v_f)$
            \label{line:stackfallback}
            \State $v \gets v_f$
            \label{line:indirectlast}
            
        \EndIf
        
        \State $r \gets r -2$
       \EndWhile

        \State $\emph{committedRound} \gets v.round$ 
        \State $order\_vertices()$
            
    \EndProcedure

        \end{multicols}
    \vspace{-2em}
    
\end{algorithmic}
\label{alg:modes}
\end{algorithm*}

\end{document}